# AI-Assisted Model for Generating Multiple-Choice Questions

Tetiana Krushynska, Jani Ursin, Ville Heilala

University of Jyväskylä, Finland

**Abstract**

Multiple-choice questions (MCQs) are widely used across diverse educational fields and levels. Well-designed MCQs should evaluate knowledge application in real-world situations. However, writing such test items in sufficient numbers is challenging and time-consuming, especially in natural science education. The problem of a sufficient number of MCQs has two aspects: content coverage and exam security. Therefore, generating test items involves two tasks: creating MCQ prototypes and transforming these prototypes into item series. In automated item generation, prototype creation aligns with template-based methods like cognitive modelling, while item expansion corresponds to example-based techniques. The aim of this research was designing the goal-oriented conceptual model of human–AI co-creation of MCQs that should meet strictly formulated quality criteria. The resulting three-step model for creating MCQ prototypes distributed prompts between several AIs, with human revision of responses for each prompt before setting the next one. To transform the MCQ prototype into an MCQ series, a one-step model was developed in which multiple new items are generated simultaneously. These items assess the same learning outcome but are not simple rephrasings of the prototype or of one another. Based on human and automated evaluation, approximately half of the output MCQs were acceptable without editing. Minor corrections of initially rejected test items allowed for a moderate increase in acceptance of MCQs in series and a significant improvement of MCQ-prototypes.

## Introduction

Tests with multiple-choice questions (MCQs) are widely used in various fields and levels of education in many countries. The standardization of test items, which covers both questions and answers, allows for formalized and even automated scoring. This reduces subjective errors (Singh & Villiers, 2012) and makes measuring learning outcomes effective, flexible, and inclusive (Tawalare et al., 2020). Another benefit of MCQ-based assessment is its compatibility with learning management systems (Babo & Suhonen, 2020). The problem lies in sufficient quality and quantity of test items. The writing of new MCQs is challenging and time-consuming (Kumar et al., 2024), indicating that improvements in this process would meaningfully address the practical needs of educators.

Advances in technology are increasingly positioning Large Learning Models (LLMs) as a powerful tool for educators in the development of various types of educational materials, including explanatory and instructional texts, scenarios of learning activities, assignments, and quizzes containing different kinds of tasks (e.g., Lan et al., 2026). Exposure to AI-generated suggestions contributes to real-world connections that make content immersive and interdisciplinary. In grading students, artificial intelligence (AI) can support in creating the assessment rubrics and directly generate tests based on standardized content (Kshetri, 2023). Automatic question-generation techniques help test developers to construct large numbers of high-quality questions (Kurdi et al., 2020). In teacher-AI interaction, AI play a helpful role in idea generation but that ideas require human refining and validating (Chen, 2024). The main patterns of teacher – AI interaction comprise sharing and comparing information, discovering issues and exploring differences, negotiating meaning and knowledge construction, testing and modifying plans, and consensus declaration. Such behaviour patterns emerge in a dynamic loop where teachers actively guide the generative AI, critically assess its contributions, and refine outputs to achieve optimal educational materials (Li & Jiang, 2025).

This study aims to develop a conceptual model for AI-assisted multiple-choice test item generation that integrates commonly available large language model tools without requiring specialized IT competencies, while emphasizing the essential role of subject-matter and pedagogical expertise. The model positions AI as a supportive instrument that enhances the productivity of MCQ developers and reduces educators' workload, rather than as a substitute for human judgment in test-item design.

**Background**

*Requirements for MCQs' quality and quantity*
From the point of view of goal-oriented conceptual modelling, the main goal of obtaining a superior set of test items (item bank) can be decomposed into several sub-goals that reflect different aspects of the quality and quantity of generated MCQs. Educators quite often refer to MCQs as primarily effective for evaluating factual knowledge based on memorization (Haladyna, 2022). The orientation toward evaluating higher-order thinking skills can also be possible (Liu et al., 2023), but it sets specific requirements for the entire MCQ and its structural elements. Like any subject-specific assessment tasks, MCQs should align with the expected learning outcomes of the target test-takers and be scientifically and linguistically correct. Being a 'closed question', the MCQ should be framed in a way that provides sufficient and self-contained information, enabling students of a defined specialty and educational level to identify the correct answer based on the information presented (Worley, 2015). Well-designed MCQs have a relatively long stem that presents a realistic, context-rich scenario. Such a scenario exemplifies a specific, plausible instance of a broader scientific principle or pattern in a way that is new to the student, not covered in class, and thus subject to recall (Brookhart, 2010). The stem describes a situation that could occur in a relevant scientific or professional context, along with appropriate actions, settings, and roles. For evaluating higher-order thinking skills, a clear logical connection between the stem and the question is necessary. It should be impossible to answer the question correctly using prior knowledge alone, without analysing the details provided in the stem. The stem does not merely introduce a topic, but provides necessary data to the reasoning process, resulting in the selection of the correct answer.

Among answer options, only one must be definitively correct while other options are incorrect, incomplete, or less appropriate, based on the information provided in the stem. To prevent guessing behaviour, it is essential that the stem and the correct answer do not share exact or closely related key terms (Ibbett & Wheldon, 2016), unless those terms are also represented in one or more distractors. The distractors should not be clearly irrelevant or based on concepts not supported by the text of the stem. Required homogeneity of answer options means that distractors possess lexical, semantic, and syntactic similarity; they are related to the same domain or framework as the correct answer and appear equally attractive to students.

The goal-oriented conceptual model for generating MCQs should consider the quality requirements mentioned above as sub-goals. Following such complicated requirements makes creating MCQs, oriented to higher-order thinking skills, challenging and time-consuming (Kar et al., 2015). The involvement of LLMs can be beneficial in this regard due to their ability to analyse vast amounts of data from various sources to identify current trends and real-world problems withing defined context. That ensures MCQs content is relevant, up-to-date, and oriented to practical applications of theoretical concepts in real life.

The fact that creating even a single MCQ is very time-consuming draws attention to the question of how many test items we need. The problem of the necessary number of MCQs has two aspects: 1) content coverage and 2) exam security and integrity. The content coverage requires the set of test items to evaluate outcomes related to each didactic unit (concept) of the learning course (Koller et al., 2017). These MCQs should differ conceptually. They share "stem - question – correct answer – distractors" structure, while their contextual and linguistic difference/similarity is not essential. The number of such test items depends on the number of concepts to be assessed for acquisition. After the item bank is completed, the need for new MCQs arises when the educational content changes.

Exam security and integrity is higher when multiple-pool designs used in an item bank construction (Zhang et al., 2012). This approach allows for dividing MCQs into 'open' and 'secret' item banks – for training (formative assessment) and evaluation (summative assessment), respectively. The MCQs that occupy the same places in the structure of both item banks would be conceptually identical but differ in their contexts and wording. Moreover, in the 'secret' item bank, it should be a series of MCQs corresponding to each of the example MCQs in the 'open' item bank. This would avert memorization of the MCQs from the 'open' item bank as a strategy of preparing for the exam but encourage students to meaningful study of learning content. The 'secret' item bank should be big enough to ensure compiling numerous variants of an exam test, where formulations of test items are different across the variants, but they evaluate the same learning outcomes and have equal complexity. The necessary number of test items in an individual series depends on the number of test-takers or the required number of test variants. Considering that the reuse of test items leads to their leakage and affects exam scores (Wagner-Menghin et al., 2013), the need for new MCQ variations arises after each testing cycle.

The connection between two aspects of the necessary number of MCQs is that items, created with the aim of content coverage, may serve as prototypes for creating MCQ series aimed at maintaining exam security and reliability. This approach divides the general task of providing a sufficient amount of test items into 1) creating MCQ-prototypes and 2) transforming prototypes into series. The essential sub-goal in generating MCQ prototypes is their alignment with learning content and objectives, including higher-order thinking skills and knowledge application in real-world situations. For generating a series of MCQs, the key point is that their similarity to the prototype and to each other should not be obvious to test-takers. Thus, models of AI-assisted generation of MCQ prototypes and MCQ series will differ significantly.

*Current approaches for AI-assisted generating MCQs*

Methods of automatically generating test items can be primarily categorized into template-based and NLP (natural language processing)-based. These methods differ in the form of input given to the AI and in the balance of human and AI involvement (Gorgun & Bulut, 2025). In the template-based approach, the input is an item model, manually developed by a subject-matter expert using cognitive modelling. The item model includes variable elements with specified options, which AI manipulates to generate test items. This often allows for creating many items with adequate didactic and psychometric quality (Falcão et al., 2024). However, a similar model, where subject-matter experts matched concepts within educational content and specified the question's types, produced MCQs oriented to simple memorization (Witchel et al., 2018).

In contrast, the NLP-based approach typically uses educational texts or examples of test items as input. It may also accept explicitly stated learning outcomes as input for generating test items. Here, NLP techniques leverage structural, semantic, and contextual analysis to convert texts into questions (Kumar et al., 2018). MCQs could be generated by defining key words in the subject-matter text, selecting sentences containing these words, replacing the key word with a gap or an appropriate wh-word to obtain a question (stem), and finally adding distractors taken from the whole body or clusters of keywords (Nwafor & Onyenwe, 2021; Das et al., 2021). However, the formal text analysis, which does not consider the specificity of the learning discipline, often resulted in producing factual MCQs oriented to low-order thinking. Conversely, this approach performs better in crafting reading comprehension questions (Huang & He, 2016). Notably, asking an LLM directly to create MCQs according to provided examples allowed for generating test items mainly acceptable, containing clinical scenarios and requiring higher-order thinking skills (Klang et al., 2023). When an LLM is provided with examples (initial text + test item, based on that text), it can produce similar items, with a defined level of complexity withing certain context (Maity et al, 2025). But MCQs, obtained in that way, rarely contained real-world details and so failed in assessing the skills in knowledge application (Agarwal et al., 2023).

Since both mentioned approaches can be effective in generating MCQs, but no one of them can guarantee the quality of the created items, a hybrid approach, aimed to improve the template-based method through various NLP techniques, was suggested (Shin, 2021). Importantly, in all cases, experts' review of the clarity, accuracy, and relevance of AI-generated items was necessary. The creation of high-quality MCQs required multiple revisions of prompts to refine distractors, context, difficulty, or the conceptual focus (Law et al., 2025).

In the frameworks of goal-oriented conceptual modelling, AI techniques, algorithms, and educator review processes serve as means to reach the goal of generating assessment tasks of sufficient quality and quantity. From the perspective of two approaches in automated creation of test items, the task of transforming MCQ-prototypes into series matches the example-based NLP techniques, which may involve operations such as paraphrasing, analogy, substitution, inversion, negation, formalization or humanization, condensation or expansion, and transformation of textual content into visual formats. While creating MCQ-prototypes seems to be aligning with the template-based method involving cognitive modelling. It should be noted that the template-based approach, on its own, cannot reduce the human workload because templates are created manually. Deep and precise knowledge, which subject-matter experts have within learning content, is necessary to present the logical structure of that content as a hierarchical semantic network or concept map. Experts are also familiar with the educational standards: in the general structure of a learning subject, they can identify concepts that match one or another learning outcome. Such elements of the semantic network serve for building the template of the test item, aimed at evaluating the corresponding learning outcome. The hybrid approach employing NLP techniques could streamline this process: LLMs are able to analyse the text of a subject textbook to find its main concepts and logically organise them into a concept map (Yang et al., 2025). Further transformation of defined concepts into a complete test item may include intermediate steps, like formulation of a related real-world problem + decision, and adding appropriate distractors. The result will be as successful as clear the goal-oriented model describes the desirable features of MCQs.

*Tools for AI-assisted generation MCQs*

Over the last 10-15 years, there were educational studies that included the development of original MCQ-generators with their own interface. They reported that educators estimated such applications positively, and sometimes the apps were implemented at a university or department level (Falcão et al., 2024; Pugh et al., 2020). However despite increasing experimental research, there is no publicly available reports indicate that national education authorities or professional licensing bodies have officially adopted AI systems for autonomous test-item generation in high-stakes examinations. These bodies typically rely on expert committees for rigorous validation of test items. Primary generating MCQs is usually performed through the involvement of subject-matter teachers who write new items according to provided guidelines and templates. These 'raw' MCQs are then revised by experts. High acceptance standards imposed by the testing board necessarily limit the proportion of accepted test items that causes ineffective workload on university teachers.

At the same time, LLMs like ChatGPT became freely available and widely used. The primary function of LLMs lies in their capacity to produce contextually relevant textual outputs that support a wide range of interpretive and transformative tasks, performing the analysis, synthesis, and summarization of information. Rather than merely generating text, LLMs facilitate the reconfiguration of existing content into new forms that align with specific goals, including educational ones (Xing et al., 2025). General purpose LLMs such as ChatGPT are already equipped with generalizable linguistic and conceptual knowledge. Without the necessity of additional training or fine-tuning, educators can use them through prompt-based interactions to create content tailored to different learning needs and contexts (Lendvai, 2025).

Unlike other educational texts, MCQs have a specific "stem + prompt + answer options" structure. Nevertheless, they are natural language texts, which is the kind of output produced by LLMs. Freely available generative AI systems are designed for general purposes; therefore, teachers' pedagogical and subject-matter expertise is crucial for human–AI co-creation of high-quality MCQs (Law et al., 2025). On the other side, the involvement of generative AI in evaluation procedures, especially high-stakes exams, brings potential risks and raises ethical concerns. That might be the main reason for official examination bodies to restrict AI implementation for the purpose of generating exam tasks.

*Legal and ethical aspects of AI-assisted generation of test items*
According to the AI Act adopted by the European Union (2024), implementing AI systems in education should be classified as a high-risk factor if they are used for determining access or admission and materially influence the level of education that individuals will receive. The risk mainly relates to evaluating learning outcomes of persons and monitoring students' behaviour during tests (Regulation (EU) 2024/1689, Annex III; European Union, 2024). However, this document specifies exceptions to being high-risk AI, among which there are the AI systems intended to perform a preparatory task to an assessment (Regulation (EU) 2024/1689, Article 6(3); European Union, 2024). The development of test items can be considered a preparatory task, as AI first analyses or transforms text and then generates MCQs, which are subsequently reviewed by humans before being used to evaluate students' learning outcomes. Thus, AI is not involved in evaluation procedures and does not influence the decisions about students' scores.

The AI Act does not object to using free and open-source AI systems/models, including for educational purposes, unless they are placed on the market or put into service as high-risk or prohibited (Regulation (EU) 2024/1689, Article 2; European Union, 2024). When assessment tasks or any other educational content is generated with the involvement of AI, which substantially alters the input provided, it falls under the Transparency obligations (Regulation (EU) 2024/1689, Article 50; European Union, 2024), i.e., should be marked as artificially generated or manipulated.

Uploading texts to AI for generating MCQs on their basis also raises copyright issues. Printed handbooks and their digital versions often explicitly mandate the prior written permission of the publisher for reproduction, distribution, and storage in a database of any part of the publication in any form or by any means. In this case, uploading even a small fragment to an AI tool may technically be a reproduction, which is restricted by copyright notice.

In contrast, for textbooks which are under the Creative Commons license, use, sharing, adaptation, distribution, and reproduction in any medium or format is permitted, while it is necessary to give appropriate credit to the original authors and the source, and indicate if changes were made. AI-assisted creation of MCQs aligns with the fair use of handbook excerpts because it is performed for educational purposes, includes non-commercial analysis and transformation of the text, and doesn't affect the market value of the original work. The concern about the use of uploaded content for AI training (Pearson, 2025) can be addressed through activation of the corresponding opt-out options to disable training and storage. Moreover, often it is possible to provide AI with a link to an open-access text instead of uploading this text in Word or PDF format, which eliminates the question of text reproduction.

Educational standards, programs of national exams, and other official documents that formulate learning objectives can also be used as input for AI-powered generation of test items. The EU aims to harmonize copyright rules and to introduce mandatory exceptions for educational uses (European Union, 2019), but the member states often retain copyright over their official publications, and the use of these documents is covered by the same regulations as the use of other original texts. The exceptions are a content of EU legislation published in the Official Journal of the European Union, all Ukrainian official documents, as well as most U.S. federal government works, which are not subject to copyright.

**Research questions**
1. This study seeks to answer the following research questions:How can a teacher – AI interaction be designed to generate high-quality MCQ prototypes for assessment in natural science education?
2. How to employ LLM for transforming the MCQ prototype into an MCQ series, providing an optimal balance of shared and distinctive features across test items?

Each of the research questions addresses the model's effectiveness under simulated conditions (a trade-off between quality and quantity of generated MCQs) and its limitations that iterative refinement, based on simulation results, would not be able to overcome.

**Research design and methods**

*A protocol for MCQ generation*

Employing the goal-oriented conceptual modelling to design human–AI co-creation of MCQs emphasizes the desired outcome as obtaining reliable assessment tasks (main goal), i.e., test items which meet strictly formulated quality criteria (sub-goals). Mapping goals to means was performed through involving several AI tools together with the prompt engineering aimed to lead MCQ developers toward a defined outcome. The modelling included several iterations based on simulation results.

Preliminary one-step model included prompt: 'Write multiple-choice question for students of [*defined level of education, speciality, and learning discipline*] oriented to higher-order thinking, with the stem formulated as a real-world situation and five answer options, among which only one is correct. Keep academic style and avoid unnecessary details. Provide the explanation why one answer is correct, and the other options are not correct.'

The prompt was applied under multiple combinations of modelling conditions, each comprising several variations to capture a broad range of settings. (Tab.1).

Table 1. Simulation conditions

| Condition | Variations |
|---|---|
| AI | MS Copilot (GPT-4), ChatGPT (GPT-4.1), DeepAI (GPT-4), Wrizzle (GPT-4o mini). All available in May – August 2025. |
| Input type: 1. Textbook fragments (**TB**) | Chapters or paragraphs from open access biology textbook for university students (Clark et al, 2018; Bruslind, 2019; Bartee et al, 2017; Smith & Selby, 2017), from 3 to 20 pages. Selected topics represented biochemistry, immunology, evolution, biotechnology, microbiology, and cell biology. |
| 2. Learning objectives (**LO**) | Formulations were taken from educational documents (Ministry of Education and Science of Ukraine, 2020; American Society for Microbiology, 2024; Kushner et al., 2022). They were related to university courses of general biology, microbiology, and virology, and biology section of matriculation exam. |
| Distribution of topics | The same topic (**1t**) for all AIs<br>Four different topics (**4t**) for four AIs |
| Required number of generated MCQs | Singl test item<br>Five test items |

The aim of this step was to reveal the most common flaws in generated MCQs for clear formulation of quality criteria of acceptable test items, and collect reusable adjustment prompts that help to meet those criteria.

At the next step of modelling, the sequence of three prompts was used: 1. develop the concept map (hierarchical semantic network) of some piece of educational content; 2. write the question with a real-world scenario which require higher-order thinking and provide correct one-phrase answer; 3. generate complete MCQ which should meet quality criteria provided (see an example in Appendix A). This multi-step model aimed to reveal which step is performed better across the tested AIs. The final model for generating MCQs optimally distributed three prompts, mentioned above, among four AIs. It also considered human control over the output of each step (Fig.1).

The model for transforming an MCQ prototype into a series performed this task directly in response to the prompt to generate 5 MCQs that differ from the provided example and from each other as much as possible, while focusing on the same problem. Another simulation began with analysing the MCQ prototype to identify the underlying concept or competence, after which 5 MCQs related to the same concept were generated. In both cases, quality criteria were applied.

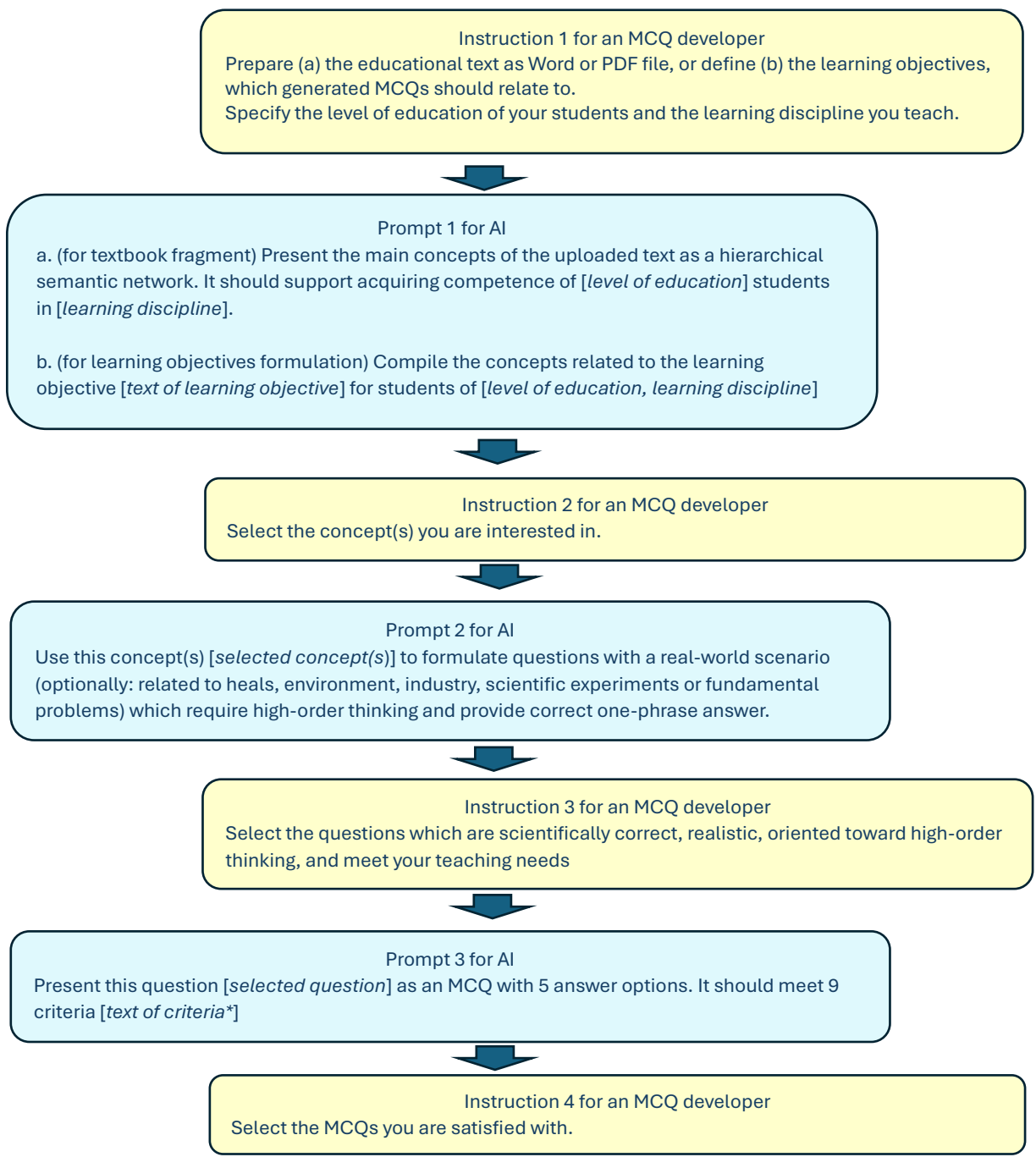

**Fig.1** Model of generating MCQ-prototypes
\* The text of the criteria is provided in Tab.2.

*MCQ Quality Assessment*
The evaluation of the quality of produced test items against nine criteria, related both to formal and intrinsic features of properly designed MCQs, was performed manually and automatically. To assess the level of agreement between human and AI's evaluations of MCQs, Cohen's kappa coefficient ($\kappa$) was used (Cohen, 1960; Sabharwal, 2021):

$$\kappa = \frac{P_o - P_e}{1 - P_e}$$

$P_o$ – the observed proportion of agreement;
$P_e$ – the expected agreement by chance, derived from the marginal totals of the contingency table.

Table 2. Contingency of human and AI decisions about acceptability of MCQs

|  | AI evaluation: Yes | AI evaluation: No | Row Total |
|---|---|---|---|
| Human evaluation: Yes | a | b | a+b |
| Human evaluation: No | c | d | c+d |
| Column Total | a+c | b+d | N |

$$P_o = \frac{a + b}{N}$$

$$P_e = \frac{(a + b)(a + c) + (c + d)(b + d)}{N^2}$$

$N$ – total number of evaluated MCQs
Interpretation of $\kappa$ values followed conventional benchmarks:
    0.01–0.20: Slight
    0.21–0.40: Fair
    0.41–0.60: Moderate
    0.61–0.80: Substantial
    0.81–1.00: Almost perfect

The conceptual similarity of generated MCQs was detected using NLP technics, i.e. setting the corresponding two subsequent prompts to MS Copilot:
*Prompt1.* In the file uploaded, what are the main concepts of MCQ1?
*Prompt2.* Do MCQs2-10 focus on the same concepts that MCQ1? You should compare the main ideas presented in MCQs, i.e., do they evaluate the same or different piece of knowledge/ competence.
The contextual and linguistic similarities between prototype and variations within the MCQ series were measured according to Tversky's Feature Model. This model connects human perception of similarity/ difference of two texts and the formal features of those texts. Tversky's model compares two texts based on common features, shared between both items, and distinctive features, unique to each item. The similarity between the two test items A and B was calculated as:

$$S(A, B) = \theta (A \cap B) - \alpha (A - B) - \beta (B - A)$$

Where:
    A∩B number of features presented it both MCQs;
    A−B number of features presented it MCQ A but absent in MCQ B;
    B−A number of features presented it MCQ B but absent in MCQ A;
    θ, α, β are parameters that control the importance of common vs. distinctive features.
If "S" is negative, readers would likely perceive two compared texts as different.
The values assigned to these parameters were θ = 1, α = β = 0,5 because, in comparing the texts, people tend to prioritize similarity to differences (Gati & Tversky, 1984).
We conducted a pairwise comparison of the MCQs within one series with the MCQ prototype and with each other. Measuring contextual similarity first required identifying the contexts represented in the items. Linguistic similarity was calculated based on the number of shared and unique words in the test items being compared (see an example in Appendix B). The originality of generated

MCQs was checked with the help of a plagiarism detector, Turnitin Draft Coach (Turnitin, LLC, Oakland, CA, USA). The threshold for considering MCQ to be original was set as less than 10% overlap with any other texts.

**Findings**

*Generating MCQ-prototype*

In 16 cycles of simulation, performed with direct prompt to generate one and then five MCQs, a total of 96 test items were produced. Of these, 36 items (37.5%) were of sufficient quality (Table 3). The effectiveness varied across used AIs from 58,33 % (ChatGPT) to 20,83% (DeepAI). In this model, formulating learning objectives yielded a higher proportion of acceptable MCQs than uploading textbook fragments (47.91% vs. 27.08%).

Table 3. Number of acceptable MCQs in response to direct prompt for a single AI

| Input* | MS Copilot | | ChatGPT | | DeepAI | | Wrizzle | | totally |
|---|---|---|---|---|---|---|---|---|---|
| | 1 MCQ | 5MCQs | 1 MCQ | 5MCQs | 1 MCQ | 5MCQs | 1 MCQ | 5MCQs | |
| 1t-TB | 0 | 0 | 1 | 2 | 1 | 0 | 0 | 1 | 5 |
| 1t-LO | 0 | 5 | 1 | 4 | 0 | 2 | 1 | 4 | 17 |
| 4t-TB | 0 | 3 | 1 | 2 | 0 | 0 | 0 | 2 | 8 |
| 4t-LO | 0 | 0 | 1 | 2 | 0 | 2 | 0 | 1 | 6 |
| totally | 0 | 8 | 4 | 10 | 1 | 4 | 1 | 8 | **36** |

*See Input specification in the "Research design and methods" section, Tab.1.

Analysis of defects in the rejected MCQs allowed for summarizing the most common flaws. Setting the adjustment prompts increased the acceptance of MCQs for 52,08%. The effective prompts were integrated into quality criteria, applied for further models (Tab. 4).

Table 4. Quality criteria based on the most common flaws revealed in generated MCQs

| Flaws | Quality criterion addresses flaws |
|---|---|
| • MCQ is incorrect scientifically<br>• Essential data are missed, making it impossible (for students of defined level of education) to select correct answer | **1**. All information presented in the MCQ — including the stem, correct answer, and distractors — must be fully scientifically accurate (every phrase separately and entire text) and can be supported by references, such as scientific articles, books, or textbooks, It is unambiguous, and free from misleading simplifications, assumptions, or comparative claims that could distort understanding.<br>This includes:<br>Avoiding oversimplified generalizations (e.g., implying that only one group possesses a function that is actually shared).<br>Ensuring that comparative or superlative terms (e.g., "better," "more," "only") are used only when clearly supported by evidence and contextually valid.<br>Verifying that the correct answer is not just more correct than the others, but entirely correct on its own merits and in alignment with scientific consensus. |
| • The stem is too short to provide enough contextual details.<br>• Formulations of answer options are too long and complicated | **2**. The MCQ must have:<br>• a long stem containing at least 3 sentences,<br>• 5 short answer options each containing no more than 7 words. |
| • Some details or situation as whole are not realistic. | **3**.The stem of the MCQ presents a realistic, context-rich scenario that exemplifies a specific, plausible instance of a broader scientific principle or pattern (for example, subject-matter pattern: 'bacteriophages can selectively kill bacteria', real-world scenario: a case of phage therapy of a certain bacterial |
| • No actual connections between subject-matter content and real-world situation. | infection, with necessary realistic details). Scenario must describe a plausible situation that could realistically occur in the relevant scientific or professional context. All actions, settings, and roles must be scientifically appropriate and consistent with how such work is typically conducted in real life or clearly framed as a model or simulation if not. |
| • MCQ does not fit level of education or educational content or learning objective. | **4.** The entire text of the MCQ must provide a complete and self-contained set of information that allows students of a defined speciality and level of education to make a conscious decision about the correct answer. All concepts, terms, and procedures referenced in the stem must be clearly stated and should align with the expected knowledge base of the target learners (for example, the MCQ for an introductory biology course can be based on a medical-related scenario, but it shouldn't refer to specific medical knowledge). The question should not rely on inference, unstated assumptions, or ambiguous phrasing that could confuse students or require knowledge beyond the course scope. |
| • No close connection between stem and question: it is possible to answer the question separately, not looking at the stem | **5.** There must be a clear and necessary logical connection between the stem and the question. The stem must provide specific information—such as data, context, or observations—that is essential for selecting the correct answer. It should be impossible to answer the question correctly using prior knowledge alone, without analyzing the details provided in the stem. The stem shouldn't merely introduce a topic, it must contribute necessary information to the reasoning process. |
| • Question is oriented to recall, not to understanding, analysis of provided information, or other thinking procedures. | **6.** MCQ requires higher-order thinking to select the correct answer, not just recall some fact. (NOT like in this example: 'In the context of antibiotic resistance, if a hospital-acquired infection presents with a mutation arising through non-replicative transposition of a transposon, what key feature distinguishes this mechanism from replicative transposition?<br>  A. The transposon remains in the original location after the process (correct answer)<br>  B. The process involves the replication of DNA at the insertion site.<br>  C. There is an integration of mobile genetic elements without any DNA breakage.<br>  D. It creates a copy of the transposon at a new site without altering the original.<br>  E. The transposase enzyme is not involved in the mechanism.' |
| • More than 1 correct answer.<br>• The 'correct' answer is actually incorrect. | **7.** Only one answer option must be clearly and definitively correct based on the information provided in the stem. The stem must include sufficient detail to rule out all other options as incorrect, incomplete, or less appropriate. If more than one option could reasonably apply without additional clarification, the criterion is not met |
| • Obviously incorrect distractors.<br>• Irrelevant distractors. | **8.** All distractors must be conceptually related to the same domain or framework as the correct answer and be plausible responses based on the scenario. They should require a comparable level of reasoning and appear equally relevant or attractive to a student who does not yet know the correct answer. There should be no distractors that are clearly irrelevant, implausible, or based on concepts not supported or implied by the information in the stem |
| • Direct hint: the same or very similar wording in the stem and the correct answer. | **9.** The stem and the correct answer should not share exact or closely related key terms—especially those central to the question's logic—unless those terms are also represented in one or more distractors. A term is considered "closely related" if:<br>• Exact word matches, including abbreviations, acronyms, and sharing the same root or meaning (e.g., "adhesins" vs "adhesion factors"),<br>• Synonyms or paraphrases that provide semantic and conceptual overlap, even if wording differ (e.g., "biofilm formation" vs "quorum sensing-dependent attachment"),<br>• It would be recognized by a test-wise student as semantically linked based on the context provided in the stem. |

| | Shared general terms (e.g., "cell," "bacteria") are acceptable, but repetition of unique terminology or context-specific words (e.g., "nutrient," "porin," "mutation", "resistance") should be avoided if it makes the correct answer more salient than the others. |
|---|---|

The model incorporating a sequence of three prompts (see Fig. 1) with integrated quality criteria was applied under the same simulation conditions as the preliminary model and appeared to be more effective (Tab.5). Forty-six MCQs (47,91%) from 96 generated ones were acceptable.

Table 5. Number of acceptable MCQs generated by multi-step model

| Input* | MS Copilot | | ChatGPT | | DeepAI | | Wrizzle | | totally |
|---|---|---|---|---|---|---|---|---|---|
| | 1 MCQ | 5MCQs | 1 MCQ | 5MCQs | 1 MCQ | 5MCQs | 1 MCQ | 5MCQs | |
| 1t-TB | 1 | 2 | 1 | 2 | 0 | 1 | 0 | 2 | 9 |
| 1t-LO | 1 | 3 | 1 | 3 | 0 | 4 | 1 | 0 | 13 |
| 4t-TB | 0 | 2 | 0 | 3 | 1 | 4 | 0 | 3 | 13 |
| 4t-LO | 0 | 2 | 1 | 4 | 0 | 2 | 0 | 2 | 11 |
| totally | 2 | 9 | 3 | 12 | 1 | 11 | 1 | 7 | **46** |

*See Input specification in the "Research design and methods" section, Tab.1.

There was no essential difference between using textbook fragments or learning objectives in the initial prompt. Both types of input were revealed as applicable. Comparing AIs' response to prompts 1- 3 showed that ChatGPT compiled the best concept maps, MS Copilot formulated the best real-world questions, while at the final stage, i.e., generating complete MCQ, all AIs performed well. It should be noted that Wrizzle also produced rather good concept maps and real-world questions when the input was learning objectives.

The sequence of the prompts was distributed among four AIs. Firstly, ChatGPT compiled a concept map, then, based on selected concepts, MS Copilot created real-world questions with correct answers, and finally, all 4 AIs used these questions to generate complete MCQs. The simulation was performed with five example topics for each type of input. Involving multiple AIs gave better results than if a single AI was used (Tab.6).

Table 6. Number of acceptable MCQs generated when prompts were distributed among tested AIs

| AIs (response to final prompt) | Input | |
|---|---|---|
| | Textbook fragment | Learning objectives |
| MS Copilot | 4 | 2 |
| ChatGPT | 3 | 4 |
| DeepAI | 2 | 3 |
| Wrizzle | 2 | 2 |
| totally | 11 from 20 - 55% | 11 from 20 - 55% |

The primary reason for rejecting MCQs was failure to meet criteria 8 and 9, namely the presence of obviously incorrect or irrelevant distractors or the inclusion of a direct hint caused by wording similarity between the question and the correct answer. In many cases, such MCQs could be improved by adding a targeted prompt instructing the model to revise the item in accordance with the relevant criterion. In other cases, the most efficient correction involved manually modifying one to five words in the MCQ text. These adjustments increased the acceptance rate of MCQs generated from textbook input to 70% and of those based on learning objectives to 85%.

The acceptance of generated MCQs, shown in Tables 3, 4, and 6, reflects human evaluation. It was compared with two automated evaluations against the same 9 quality criteria. The evaluation model contained a prompt to check whether uploaded MCQs meet the quality criteria. Eighteen best (meet all 9 criteria) and 40 worst (failed in several criteria) MCQs were selected to test the model.

In the evaluation of the 40 lowest-quality MCQs, MS Copilot identified only 22 as failing to meet the quality criteria, whereas ChatGPT detected flaws in 35 MCQs, including all 22 items rejected by Copilot. The calculated values of Cohen's kappa showed that the evaluation model provides moderate human – AI agreement regarding the acceptability of generated MCQs. (Tab.7).

Table 7. Agreement of human and AIs' decisions about accepting MCQs.

|  | AI evaluation: Yes | AI evaluation: No | Row Total | Cohen's kappa ($\kappa$) |
|---|---|---|---|---|
| MS Copilot | | | | |
| Human evaluation: Yes | 18 | 0 | 18 | 0,432* |
| Human evaluation: No | 18 | 22 | 40 | |
| Column Total | 36 | 22 | 58 | |
| ChatGPT | | | | |
| Human evaluation: Yes | 11 | 7 | 18 | 0,501* |
| Human evaluation: No | 5 | 35 | 40 | |
| Column Total | 16 | 42 | 58 | |

\* $\kappa$ value 0.41–0.60 is conventionally moderate.

Similar reasons of rejection, when at least one shared criterion was mentioned in all evaluations, were detected regarding 18 MCQs, i.e., 45% of all reviewed worst MCQs, or 81,81 % of MCQs agreed in rejection.
Six best MCQs on different topics, accepted by both humans and AIs, were selected to be used as prototypes for transformation into series.

*Generating MCQ series*

Transformation of MCQ-prototypes into series was performed according to example-based approach, by setting the direct prompt: "Generate 5 MCQs which should differ as much as possible but be focusing on the same problem and have the same difficulty as the example provided [*MCQ prototype*]. In generated MCQs, the formulation of key words in stems and correct answers should differ from that in initial MCQs. The MCQs are for [*learning discipline*] and [*level of education*]".
In another simulation, AIs firstly analysed the MCQ-prototype to reveal the concept it relates to. The prompt was: "Use this MCQ [*MCQ prototype*] as a prototype and generate 5 MCQs related to the same concept and have the same difficulty but differ on their contexts and phrasing/ wording as much as possible. The MCQs are for [*learning discipline*] and [*level of education*]".
Both simulations were repeated three times using MCQ-prototypes from three disciplines: immunology, microbiology and biochemistry; 9 quality criteria were applied. The model produced about a half of acceptable MCQs (Tab.8).

Table 8. Number of acceptable test items in generated MCQ series

| AIs | Input | | | |
|---|---|---|---|---|
|  | MCQ-prototype as an example | | Concept derived from MCQ-prototype | |
|  | acceptable | did not meet criteria* | acceptable | did not meet criteria* |
| Copilot | 8 | 4,8,9 | 10 | 1,4,5,6,8,9 |
| ChatGPT | 9 | 4,5,8 | 9 | 4,5,6,8, 9 |
| DeepAI | 3 | 2,6,8, | 6 | 6,8,9 |
| Wrizzle | 9 | 2,4,6,9 | 4 | 1,4,5,6,8,9 |
| totally | 29 from 60–48,33 % | | 29 from 60–48,33 % | |
| After minor correction | 63,33% | | 51,66 % | |

*human evaluation

After minor correction, the acceptance of example-based series rose by 15%. While MCQs, generated though deriving the concept, had more serious weaknesses. Among them, only two additional items (3,33%) could be fixed by minor correction.

Within one series, MCQ should share conceptual features. Conceptual similarity was checked by setting prompts: "What are the main concepts of MCQ1(prototype)? Do MCQs 2-10 (series) focus on the same concepts that MCQ1?"

Within one series, MCQ should present the same concept in different contexts and use different wording. Contextual and linguistic similarity was checked according to Tversky's Feature Model (see an example in Appendix B). Similarity check showed that MCQs, that belong to one series, measure learning outcomes related to the same concept, and test-takers would perceive them as somewhat similar but not identical tasks (Tab.9). When MCQ prototypes were used as direct examples, the MCQ series demonstrated more contextual diversity, while deriving concepts from MCQ-prototypes resulted in more differences in wording.

Table 9. Similarity of generated MCQs within series

| Template | Examples (1prototype + 9 variations) | Similarity of text features | | |
|---|---|---|---|---|
| | | Conceptual | Contextual | Linguistic |
| MCQ-prototype as an example | 1. Herd immunity | 100% | -4,31± 2,55 | from 16 to -36,5 |
| | 2. Peptidoglycan structure and functions | 90% | -4,49 ± 0,96 | from -18 to -38 |
| | 3. Northern blotting for measuring RNA | 100% | -3,06± 1,37 | from 8 to -26,5 |
| | average | **96,66 %** | **-3,95 ±1,63** | **from 16 to -38** |
| Concept derived from MCQ-prototype | 1. Composition and functions of capsula | 100% | -2,67± 1,73 | from -10,5 to -38,5 |
| | 2. Phospholipids and cell membrane | 90% | 0,48± 0,75 | from -3 to -35,5 |
| | 3. Mucociliary escalator as non-specific defence factor | 100% | -3,71± 1,00 | from -11,5 to -44 |
| | average | **96,66 %** | **-1,96 + 1,16** | **from -3 to -44** |

The originality check revealed that there was no borrowing MCQs from other sources. All generated test items were original, containing from 1 to 6% of similarity with other publications and databases. Mostly, shared text elements were scientific facts and subject-specific terminology.

**Discussion**

*Fine-tuning of the model of teacher – AI interaction for effective generation of MCQ prototypes*

Using a single direct prompt is a frequently employed strategy of generating MCQs (Kiyak & Emekli, 2024). Under the simulation conditions, our initial one-prompt model generated acceptable MCQs for a defined learning subject and level of education when corresponding learning objectives or textbook chapters were applied. Slightly better results were observed when learning objectives were used as input; however, these findings were based on a single example and are unlikely to be generalizable to the entire model.

Introducing two "preparatory" prompts for generating a concept map and question + answer, and human control of these intermediate outputs visibly improved the resulting MCQs. This approach is not typical in the automated generation of assessment tasks. Traditionally, subject-matter experts formulate a technical assignment for IT specialists and then make a final decision to accept or reject generated MCQs, rather than actively participate in the generation process (Hariyanto et al., 2025). Our multi-step model involves humans to proceed through the steps, which requires a comparatively higher workload for teachers. On the other hand, this model does not require any IT staff workload, making it applicable for individual use according to the teacher's needs for specific test items. The transition from a one-step to a multi-step model resulted in more than 10% increase in the number of acceptable MCQs.

The integration of different AIs (LLMs, text-to-speech, and video-generating models) for creating educational materials is usually employed when educational materials are multimodal, combining textual, visual, and audio elements. (Heilala et al., 2025). However, LLMs, which process texts solely, are based on diverse internal algorithms of text analysis and generation. The difference in their responses to the same prompt allowed for further improvement of the multi-step model by setting each prompt to the AI that performed it the best. In this way, an additional 8% increase in the quality of MCQs was achieved automatically, and by 22–37% following minor manual corrections. The simulations conducted showed no technical difficulties in switching between the four AI tools.

In ensuring the quality of MCQs, the critical role belongs to the quality criteria. In our study, the same nine criteria were integrated with the prompt for generating test items and used in their evaluation. Establishing criteria which guide experts' or automated evaluation of test items (Gierl et al., 2017), and providing briefly formulated quality requirements for their generation (Kiyak & Emekli, 2024) is quite common. Nevertheless, integration of a system of detailed criteria into prompts was challenging and required numerous iterations before AIs responded satisfactorily.

***Similarity and difference of transformed MCQs within series and model's options of generating multiple test items simultaneously***

In the development of a structured item bank, categorizing and labelling of MCQs is an essential prerequisite for streamlined test administration, especially for the automated compiling of tests (Dadey, 2022). The structure of the item bank reflects the structure of educational content, the architecture of the main concepts of the learning discipline. A series of test items reflects various contexts and presentations of the main concepts from educational content. This adds one more level to the structure of the item bank. Our example-based model for generating a series of MCQs exhibited the ability to meet this need. It provided sufficient balance between similarity and differences across MCQs in series, ensuring that those MCQs would evaluate the same learning outcomes but be perceived by test-takers as distinct tasks.

Beyond the example-based model, other simulations likewise led to the production of multiple MCQs. In the multi-step model, the last prompt (transform the "question + answer" into MCQ) was set to four AIs, resulting in a series of four MCQs sharing the same concept. Those MCQs demonstrated significant contextual and linguistic diversity, but the series was smaller than that produced by the example-based model. In the multi-step model, it was also possible to select several components of the concept map obtained in response to prompt 1 (Fig.1), to generate several "questions + answers", and then complete MCQs based on different concepts. Due to their conceptual difference, the resulting MCQs cannot be considered a series of test items, but they contribute to content coverage.

In the one-step model, after obtaining one MCQ in response to a direct prompt, it was possible to ask AI to generate similar MCQs based on another part of the same fragment of the textbook. Usually, 2-3 of the five generated MCQs were acceptable, even if the initial MCQ had some flaws. The resulting MCQs exhibited considerable variation when compared to one another in terms of their conceptual, contextual, and linguistic similarity/difference. The human review was necessary to categorize them as possible prototypes or variations within a series. These observations suggest that using separate models for generating MCQ prototypes and test-item series is preferable.

***Quality and quantity of generated MCQs***

In all simulations, the model has generated 352 test items. Approximately half of MCQ-prototypes and MCQs in series were acceptable, suggesting moderate overall effectiveness. However, the acceptance rate hinges on the evaluation method. In our model, applying nine stringent quality criteria and discarding any items that failed to meet at least one of them predictably reduced the number of accepted MCQs. Similarly, introducing the Item Writing Flaws Evaluation Instrument, which included 13 distinct rubrics for experts' evaluation of the quality of MCQs beyond psychometric parameters, revealed that 83% of items in the exam item bank contained at least one item writing flaw (Breakall et al., 2019). While the cumulative score in three parameters: clarity, relevance, and suitability for assessing a specific category of students, used by Chauhan et al. (2025), resulted in the acceptance of two-thirds of the generated test tasks.

The main reasons for rejection of test items generated with our model were as follows.

Firstly, test items did not align with the level of education or educational content, or learning objective. These "alignment flaws" are observed in MCQs across both manual and automated generation methods. To address these issues, integrating identified flaws into institutional guidelines and implementing a checklist for evaluation of test items was suggested (Khan et al., 2025). AI-assisted generating MCQs for natural science education requires deep subject-matter knowledge, making it difficult to formulate precise prompts to explain to AI what you need. The leverage of the official examination content framework (Yao et al., 2025) and iterative design cycles based on continuous user feedback (Sagadraca et al., 2024) are expected to mitigate this problem. Our model involved official content frameworks (i.e., learning outcomes and textbook fragments), but continuous feedback conflicted with its objectives of ensuring easy generation of MCQs.

Secondly, generating the text of MCQs, AIs tended to include details that are non-realistic from the subject-matter point of view, impossible, or unusual in the defined context. AI can generate texts when available real-world data have limitations, although in filling those gaps, it does not care about the realism of the details (Ahmad et al., 2024). To improve the factual accuracy of AI-generated texts and reduce hallucinations, Chen et al. (2025) recommended integrating external sources of information into the generation process, which enhances the system's ability to produce factually grounded responses by supplementing its internal knowledge with current and contextually relevant data. However, this approach is poorly applicable to generating MCQs, which should present concepts from an external resource (textbook) in various contexts, different from those in the initial text. Another strategy is to prompt models to critique or reflect on their own outputs (Ji et al., 2023). This helped to reduce hallucinations in generated MCQs but required setting several prompts.

Thirdly, the wording similarity of the stem and the correct answer provided clues for test-takers, even if those words do not completely coincide. To tackle this challenge, Yao et al. (2025) integrated expert-driven prompt engineering and iterative self-critique and self-correction feedback to improve the quality of test items. Other researchers (Mathur & Suchithra, 2022), who transformed educational texts into MCQs, employed the keyword extraction tool KeyBERT to formulate the correct answer. After that, the extracted keywords were removed from the stem and served as the answer, thus reducing the likelihood of similarity. However, this approach cannot ensure that the stem is free of synonyms related to the answer keywords, which potentially offer hints to experienced test-takers.

And finally, sometimes it was possible to answer the question separately, without analysing the data presented in the MCQ's stem. Thus, those MCQs were aimed more at recall than higher-order thinking. MCQ developers recognized stem-option misalignment as a specific flaw in test item construction. The introduction of the Scalable Automatic Question Usability Evaluation Toolkit (SAQUET), an open-source tool that leverages the Item-Writing Flaws rubric, was suggested for automated detection of such flaws (Moore et al., 2024). While Schmauder et al. (2025) provided instructions for their manual correction.

The widely used decision for resolving the four problems mentioned above is iterative refinement of AI-generated test items, including the setting of self-critique and self-reflection prompts. In our model, the prompt to check an MCQ against a defined criterion and make necessary corrections to meet that criterion was somewhat effective. However, attempts to integrate self-critique into the quality criteria resulted in very complicated formulations but had no positive impact on the quality of generated test items.

In the framework of human–AI interaction, increasing the number of acceptable test items while maintaining high quality can be achieved by correcting MCQs whose initial flaws are relatively minor. Such corrections may involve additional prompt engineering (Burke, 2025) or manual edits (May et al., 2025; Law et al., 2025), both of which require time and effort. This undermines one of the key advantages of using AI for the development of educational content— the reduction of educators' workload. Assuming, that the total time to produce a single high-quality MCQ should not exceed 20 minutes, a reasonable limit for post-generation correction might be up to four adjustment prompts or manual edits of no more than ten words. This threshold was applied in our study. It reflects a design that balances automated efficiency with the requirement for human oversight.

The trade-off between quality and quantity of accepted MCQs depends on educational purposes: creating test items for high-stakes exams would prioritize quality, but if MCQs are needed for current formative assessment within a newly introduced learning course, the development of a primary item bank requires fast accumulation of MCQs, which could be improved or replaced later. In both cases, the post-edition workload would be determined by the actual need for high-quality test items.

**Limitations and suggestions for further research**

The simulation involved a limited number of LLMs. Other LLMs, as well as updated versions of those used, may produce different responses to the same prompts, potentially affecting the reproducibility of research findings.

Approximately one-half of the generated MCQs had some construction flaws. There were misalignments with the level of education or learning objective, factual inaccuracy (hallucinations), hints to experienced test-takers, and the possibility of answering the question without reading the stem. The improvements of MCQs were possible through additional prompts or manual correction, but this was carried out outside the framework of the developed model.

Conducting simulations, the first author considered the role of a university teacher who has no specific computer skills. However, real evidence about how the model meets teachers' needs in generating assessment tasks would be obtained through an empirical study: introduction of the model into teachers' practice, followed by a users' feedback survey. Considering that the observed agreement between human and automated evaluations of the quality of produced test items was moderate, the survey should be supplemented with expert evaluation of the generated MCQs to ensure alignment with quality standards and applicability for assessing higher-order thinking skills within established educational programs.

**Conclusion**

The model of interaction of a teacher and commonly available AIs can ensure generating high-quality MCQ for assessment in natural science education when three subsequent prompts were distributed between several LLMs (namely, ChatGPT compiled a concept map; MS Copilot formulated question with answer; and any of AIs like ChatGPT, MS Copilot, DeepAI, or Wrizzle created complete MCQ), and responses for each prompt were checked by MCQ developer before setting the next prompt.

The example-based model can be used to transform an MCQ prototype into an MCQ series when the prompt emphasizes that the resulting MCQs should focus on the same problem or concept but differ significantly in their contexts and wording. In the series obtained in this way, the test items assessed the application of the same element of knowledge, but test-takers would unlikely perceive such MCQs as just a rephrasing of a prototype or each other.

The integration of nine strictly formulated quality criteria into prompts, aimed at generating complete MCQs, played a critical role in both models. Applying these criteria resulted in approximately half of the output MCQs being acceptable without editing, having sufficient quality, and oriented to higher-order thinking skills. Post-generation revision and minor corrections of initially rejected test items allowed for moderate improvement of MCQs in series and a significant improvement of MCQ-prototypes, so that their acceptance rate approached that usually reported for this kind of assessment tasks written by experts.

**References**

Ahmad, M., Ho, W. Y., Goyes Robalino, A. P., & Maqsood, M. (2025). Theoretical implications of generative AI for content generation in geoinformatics training. In A. Arinushkina (Ed.), *Integration Strategies of Generative AI in Higher Education* (pp. 101-130). IGI Global Scientific Publishing. https://doi.org/10.4018/979-8-3693-5518-3.ch005

Agarwal, M., Sharma, P., & Goswami, A. (2023). Analysing the applicability of ChatGPT, Bard, and Bing to generate reasoning-based multiple-choice questions in medical physiology. *Cureus*, *15*(6), e40977. https://doi.org/10.7759/cureus.40977

American Society for Microbiology. (2024). Recommended curriculum guidelines for undergraduate microbiology education: Fundamental statements and skills. Retrieved April 7, 2025, from https://asm.org/guideline/asm-curriculum-guidelines-for-undergraduate-microb

Babo, R., Suhonen, J., & Tukiainen, M. (2020). Improving workgroup assessment with WebAVALIA: The concept, framework and first results. *Journal of Information Technology Education: Innovations in Practice, 19*, 157-184. https://doi.org/10.28945/4627

Bartee, L., Shriner, W., & Creech, C. (2017). *Principles of Biology*. Open Oregon Educational Resources. pp. 74-80, 129-148, 357-387. Retrieved March 27, 2025, from https://openoregon.pressbooks.pub/mhccmajorsbio/

Breakall, K., Randles, C. A., & Tasker, R. (2019). Development and evaluation of a rubric for diagnosing item-writing flaws in general chemistry multiple-choice assessments. *Chemistry Education Research and Practice*, *20*(1), 62–74. https://doi.org/10.1039/C8RP00262B

Brookhart, S. M. (2010). *How to assess higher-order thinking skills in your classroom*. ASCD, Alexandria, Virginia USA. 159 p.

Burke, C. M. (2025). AI-assisted exam variant generation: A human-in-the-loop framework for automatic item creation. *Education Sciences*, *15(8)*, 1029. https://doi.org/10.3390/educsci15081029

Bruslind, L. (2019). *General Microbiology*. Oregon State University, Corvallis, Or. pp.31-35, 61-67, 122-127, 150-155. Retrieved March 27, 2025, from https://open.oregonstate.education/generalmicrobiology/

Chauhan, A., Khaliq, F. & Nayak, K.R. (2025). Title: Assessing quality of scenario-based multiple-choice questions in physiology: Faculty-generated vs. ChatGPT-generated questions among Phase I medical students. *Int J Artif Intell Educ*, *35*, 2315–2344. https://doi.org/10.1007/s40593-025-00471-z

Chen, B., Zhang, Z., Langrené, N., & Zhu, S. (2025). Unleashing the potential of prompt engineering for large language models. *Patterns, 6(6)*, 101260. https://doi.org/10.1016/j.patter.2025.101260

Chen, X. (2024). Collaborative design with conversational artificial intelligence: A case study of pre-service language teachers in multidisciplinary teams. *International Journal of Technology in Teaching and Learning*, 20 (2), 95-107. https://doi.org/10.37120/ijttl.2024.20.2.01

Clark, M.A., Choi, J., & Douglas, M. (2018). *Biology 2e*. Rice University, Houston, Texas. pp. 273-295, 417-428. https://openstax.org/details/books/biology-2e

Cohen, J. (1960). A coefficient of agreement for nominal scales. *Educational and Psychological Measurement*, 20(1), 37–46. https://doi.org/10.1177/001316446002000104

Dadey, N. (2022). *Item Bank*. Routledge. https://doi.org/10.4324/9781138609877-REE146-1

Das, B., Majumder, M., Phadikar, S., & Sekh, A. A. (2021). Multiple-choice question generation with auto-generated distractors for computer-assisted educational assessment. *Multimed Tools Appl 80*, 31907–31925. https://doi.org/10.1007/s11042-021-11222-2

European Union. (2019). Directive (EU) 2019/790 of the European Parliament and of the Council of 17 April 2019 on copyright and related rights in the Digital Single Market and amending Directives 96/9/EC and 2001/29/EC. *Official Journal of the European Union*, L 130, 92–125. https://eur-lex.europa.eu/eli/dir/2019/790/oj

European Union. (2024). Regulation (EU) 2024/1689 of the European Parliament and of the Council of 13 June 2024 laying down harmonised rules on artificial intelligence (Artificial Intelligence Act) and amending certain Union legislative acts. *Official Journal of the European Union*, L 2024/1689. http://data.europa.eu/eli/reg/2024/1689/oj

Falcão, F.M.V., Pereira, D.S., Pêgo, J.M. et al. (2024) Progress is impossible without change: Implementing automatic item generation in medical knowledge progress testing. *Educ Inf Technol 29*, 4505–4530. https://doi.org/10.1007/s10639-023-12014-x

Gati, I. & Tversky, A. (1984). Weighting common and distinctive features in perceptual and conceptual judgments. *Cognitive Psychology*, *16(3),* 341-370. https://doi.org/10.1016/0010-0285(84)90013-6.

Gierl, M. J., Bulut, O., Guo, Q., & Zhang, X. (2017). Developing, analyzing, and using distractors for multiple-choice tests in education: A comprehensive review. *Review of Educational Research*, *87(6)*, 1082–1116. https://doi.org/10.3102/0034654317726529

Gorgun, G., & Bulut, O. (2025). Instruction-tuned large-language models for quality control in automatic item generation: A feasibility study. *Educational Measurement: Issues and Practice*, *44(1)*, 96-107. https://doi.org/10.1111/emip.12663

Heilala, V., Araya, R., & Hämäläinen, R. (2025). Beyond text-to-text: An overview of multimodal and generative artificial intelligence for education using topic modeling. ArXiv. https://doi.org/10.1145/3672608.3707764

Haladyna, T. (2022). Creating multiple-choice items for testing student learning. International *Journal of Assessment Tools in Education*, *9(Special Issue)*, 6-18. https://doi.org/10.21449/ijate.1196701

Hariyanto, Kristianingsih, F.X.D. & Maharani, R. (2025). Artificial intelligence in adaptive education: A systematic review of techniques for personalized learning. *Discov Educ 4*, Article 458. https://doi.org/10.1007/s44217-025-00908-6

Huang, Y., & He, L. (2016). Automatic generation of short answer questions for reading comprehension assessment. *Natural Language Engineering*, *22*(3), 457–489. https://doi.org/10.1017/S1351324915000455

Ibbett, N.L., & Wheldon, B.J. (2016). The incidence of clueing in multiple choice testbank questions in accounting: Some evidence from Australia. *The E-Journal of Business Education and Scholarship of Teaching*, *10(1)*, 20-35. https://eric.ed.gov/?id=EJ1167417

Ji, Z., Yu, T., Xu, Y., Lee, N., Ishii, E., & Fung, P. (2023). *Towards mitigating hallucination in large language models via self-reflection*. ArXiv. https://doi.org/10.48550/arXiv.2310.06271

Kar, S.S., Lakshminarayanan, S., & Mahalakshmy, T. (2015). Basic principles of constructing multiple choice questions. *Indian Journal of Community and Family Medicine, 1*, 65-69. doi:10.4103/2395-2113.251640

Khan, H. F., Qayyum, S., Beenish, H., Khan, R. A., Iltaf, S., & Faysal, L. R. (2025). Determining the alignment of assessment items with curriculum goals through document analysis by addressing identified item flaws. *BMC Medical Education, 25*, Article 200. https://doi.org/10.1186/s12909-025-06736-4

Kiyak, Y. S., & Emekli, E. (2024). ChatGPT prompts for generating multiple-choice questions in medical education and evidence on their validity: A literature review. *Postgraduate Medical Journal*, 100(1189), 858-865. https://doi.org/10.1093/postmj/qgae065

Klang, E., Portugez, S., Gross, R., Lerner, K., Brenner, A., Gilboa, M., Ortal, T., Ron, S., Robinzon, Meiri, H., & Segal, G. (2023). Advantages and pitfalls in utilizing artificial intelligence for crafting medical examinations: A medical education pilot study with GPT-4. *BMC Med Educ*, 23(1), Article 772. https://doi.org/10.1186/s12909-023-04752-w

Koller, I., Levenson, M. R., & Glück, J. (2017). What do you think you are measuring? A mixed-methods procedure for assessing the content validity of test items and theory-based scaling. *Frontiers in Psychology, 8*, 218785. https://doi.org/10.3389/fpsyg.2017.00126

Kshetri, N. (2023). The future of education: generative artificial intelligence's collaborative role with teachers. *IT Professional*, *25 (6)*, 8-12. https://doi.org/10.1109/mitp.2023.3333070.

Kumar, A. P., Nayak, A., Shenoy, M., Chaitanya, & Ghosh, K. (2024). A novel framework for the generation of multiple choice question stems using semantic and machine-learning techniques.*, Int J Artif Intell Educ, 34*, 332–375. https://doi.org/10.1007/s40593-023-00333-6

Kumar, V., Boorla, K., Meena, Y., Ramakrishnan, G., & Li, Y.-F. (2018). *Automating reading comprehension by generating question and answer pairs*. ArXiv. https://doi.org/10.48550/arXiv.1803.03664

Kurdi, G., Leo, J., Parsia, B., Sattler, U., & Al-Emari, S. (2020). A systematic review of automatic question generation for educational purposes. *Int J Artif Intell Educ,* 30, 121–204. https://doi.org/10.1007/s40593-019-00186-y

Kushner, D.B., Cline, T.D., Danthi, P., Debbink, K.M., Ferran, M.C., Flenniken, M., Johnson, D.M., Mehle, A., Morales, D.A., Mukhopadhyay, S., Parker, J.S.L., Spriggs, C.C., O'Donnell, L.A., & Magnnis, M.S. (2022). Curriculum guidelines for graduate and undergraduate virology courses. *J Virol*, 96(18), e01305-22. https://doi.org/10.1128/jvi.01305-22

Lan, Y., Li, X., Du, H., Lu, X., Gao, M., Qian, W., & Zhou, A. (2026). Survey of natural language processing for education: Taxonomy, systematic review, and future trends. IEEE Transactions on Knowledge and Data Engineering, 38(1), 659–678. https://doi.org/10.1109/tkde.2025.3621181

Law, A.K.K., So, J., Lui, C.T., Choi, Y.F., Cheung, K.H., Hung, K.K., & Graham, C.A. (2025). AI versus human-generated multiple-choice questions for medical education: a cohort study in a high-stakes examination. *BMC Medical Education*, *25*, 1-9. https://doi.org/10.1186/s12909-025-06796-6

Lendvai, G.F. (2025). ChatGPT in academic writing: A scientometric analysis of literature published between 2022 and 2023. *Journal of Empirical Research on Human Research Ethics*, *20(3)*, 131-148. https://doi.org/10.1177/15562646251350203

Li, M., & Jiang, L. (2025). Research on human-machine collaboration behavior patterns supported by generative artificial intelligence. *5th International Conference on Artificial Intelligence and Education (ICAIE),* Suzhou, China, 2025, 309–313. https://doi.org/10.1109/icaie64856.2025.11157944

Liu, Q., Wald, N., Daskon, C., & Harland, T. (2023). Multiple-choice questions (MCQs) for higher-order cognition: Perspectives of university teachers. *Innovations in Education and Teaching International*, *61*(4), 802–814. https://doi.org/10.1080/14703297.2023.2222715

Maity, S., Deroy, A., & Sarkar, S. (2025). Can large language models meet the challenge of generating school-level questions? *Computers and Education: Artificial Intelligence, 8*, Article 100370. https://doi.org/10.1016/j.caeai.2025.100370.

May, T. A., Fan, Y.K., Stone, G.E., Koskey, K.L. K., Sondergeld, C.J., Folger, T.D., Archer, J.N., Provinzano, K., & Johnson, C.C. (2025). An effectiveness study of generative artificial intelligence tools used to develop multiple-choice test items. *Education Sciences*, *15(2),* 144. https://doi.org/10.3390/educsci15020144

Ministry of Education and Science of Ukraine. (2019). Standart vyshchoi osvity Ukrainy: pershyi (bakalavrskyi) riven, haluz znan 09 "Biolohiia", spetsialnist 091 "Biolohiia" [Standard of higher education of Ukraine: First (bachelor's) level, field of knowledge 09 "Biology", specialty 091 "Biology"]. Retrieved June 23, 2025, from https://mon.gov.ua/static-objects/mon/sites/1/vishcha-osvita/zatverdzeni%20standarty/2019/11/22/091-Biolohiya-1.pdf

Ministry of Education and Science of Ukraine. (2020). Prohrama zovnishnoho nezalezhnoho otsiniuvannia rezultativ navchannia z biolohii, zdobutykh na osnovi povnoi zahalnoi serednoi osvity [Programme of external independent assessment of learning outcomes in biology acquired through complete general secondary education]. Retrieved June 23, 2025, from https://testportal.gov.ua/progbio/

Moore, S., Costello, E., Nguyen, H. A., & Stamper, J. (2024). *An automatic question usability evaluation toolkit*. ArXiv. https://arxiv.org/abs/2405.20529

Nwafor, C., & Onyenwe, I. (2021). An automated multiple-choice question generation using natural language processing techniques. *International Journal on Natural Language Computing, 10*, 1–10. https://doi.org/10.5121/ijnlc.2021.10201

Pearson S.H. (15 May 2025). Understanding CC Licenses and AI training: A legal primer. Retrieved August 12, 2025, from https://creativecommons.org/2025/05/15/understanding-cc-licenses-and-ai-training-a-legal-primer/

Pugh, D., De Champlain, A., Gierl, M., Lai, H., & Touchie, C. (2020). Can automated item generation be used to develop high quality MCQs that assess application of knowledge? *Research and Practice in Technology Enhanced Learning*, *15(1)*, 12. https://doi.org/10.1186/s41039-020-00134-8

Sabharwal, C. L. (2021). Cohen's kappa statistic and new Kappa Statistic for measuring and interpreting inter-rater agreement. *International Journal of Research in Engineering and Science*, *9(7)*, 23–28. https://www.ijres.org/papers/Volume-9/Issue-7/Series-12/D09072328.pdf

Sagadraca, C., Santos, Z., Simón, D., Tolentino, M. E., & Ventura, R. V. (2024). Cognicraft: smart exam question generation with AI and Bloom's taxonomy. *ICCE 2024: The 32nd International Conference on Computers in Education.* November 25–29, 2024, Quezon City, Philippines https://doi.org/10.58459/icce.2024.4918

Schmauder, R., Daugherty, K., Rudolph, M., Hutson, B., McCarthy, K., McCauley, S., & Eagle, D. (2025). The art and science of item writing: A review of established guidelines for multiple-choice questions. *Intersection: A Journal at the Intersection of Assessment and Learning*. https://doi.org/10.61669/001c.146079

Shin, E. (2021). Automated item generation by combining the non-template and template-based approaches to generate reading inference test items. [Doctoral dissertation, University of Alberta].

Education & Research Archive. https://ualberta.scholaris.ca/items/8eb94141-c4b5-45a6-be48-0e6a53adff39

Singh, U.G. & de Villiers, M.R. (2012). Investigating the use of different kinds of multiple-choice questions in electronic assessment (e-assessment). *Progressio*, 34(3), 125-143. https://api.semanticscholar.org/CorpusID:111004197

Smith, M., & Selby, S. (2017). *Microbiology for Allied Health Students. Biological Sciences Open Textbooks.* 15. pp. 544-596, 729-737. https://oer.galileo.usg.edu/biology-textbooks/15

Tawalare, K., Pawar, J., Tawalare, K., & Karade, R. (2020). Need of multiple choice questions (MCQs) in assessment criteria of BAMS curriculum. *Journal of Education Technology in Health Sciences*, 7, 54-57. https://doi.org/10.18231/j.jeths.2020.014

Verkhovna Rada of Ukraine. (1994). Zakon Ukrainy "Pro avtorske pravo i sumizhni prava" [Law of Ukraine on copyright and related rights] (No. 3792-XII, as amended). Retrieved from https://zakon.rada.gov.ua/laws/show/3792-12

Wagner-Menghin, M., Preusche, I., & Schmidts, M. (2013) The effects of reusing written test items: A study using the Rasch model. *International Scholarly Research Notices*, 2013 (1), 585420. https://doi.org/10.1155/2013/585420

Witchel, H.J., Guppy, J.H., & Smith, C.F. (2018). The self-assessment dilemma: An open-source, ethical method using Matlab to formulate multiple-choice quiz questions for online reinforcement: Advances in physiology education. *Advances in Physiology Education*, 42(4), 697–703. https://doi.org/10.1152/advan.00081.2018

Worley, P. (2015) Open thinking, closed questioning: Two kinds of open and closed question. *Journal of Philosophy in Schools, 2(2),* 17-29. https://doi.org/10.21913/jps.v2i2.1269

Xing, W., Nixon, N., Crossley, S., Denny, P., Lan, A., Stamper, J., & Yu, Z. (2025). The use of large language models in education. *Int J Artif Intell Educ 35,* 439–443. https://doi.org/10.1007/s40593-025-00457-x

Yang, T., Ren, B., Gu, C., He, T., Ma, B., & Konomi, S. (2025). Leveraging LLMs for automated extraction and structuring of educational concepts and relationships. *Machine Learning and Knowledge Extraction*, *7*(3), 103. https://doi.org/10.3390/make7030103

Yao, Z., Parashar, A., Zhou, H., Jang, W. S., Ouyang, F., Yang, Z., & Yu, H. (2025). MCQG-SRefine: Multiple choice question generation and evaluation with iterative self-critique, correction, and comparison feedback. (Version 4). ArXiv. https://doi.org/10.48550/arXiv.2410.13191

Zhang, J., Chang, HH. & Yi, Q. (2012) Comparing single-pool and multiple-pool designs regarding test security in computerized testing. *Behav Res 44*, 742–752. https://doi.org/10.3758/s13428-011-0178-5

Appendix A.

# Example of the Model application

**Topic:** Photosynthesis and respiration
**Target category of test-takers:** upper secondary school students
**Input:** learning objective "Compare and contrast photosynthesis and cellular respiration in terms of reactants, products, energy flow, organelles involved, and ecological roles" (source - Biology section of matriculation exam, Ukraine).

Table A.1. Responses of selected AIs to subsequent prompts for generating MCQ-prototypes.

| Prompt | AI's response to prompt |
|---|---|
| 1. Compile the concepts related to the learning objective of upper secondary school biology course 'Compare and contrast photosynthesis and cellular respiration in terms of reactants, products, energy flow, organelles involved, and ecological roles' into a hierarchical semantic network. | **Chat GPT**<br>1. Biological Processes<br>• Photosynthesis<br>   o *Definition:* Process by which autotrophs convert light energy into chemical energy (glucose)<br>   o *Location:* Chloroplasts<br>   o *Organisms:* Plants, algae, some bacteria<br>• Cellular Respiration<br>   o *Definition:* Process by which cells break down glucose to produce ATP<br>   o *Location:* Mitochondria<br>   o *Organisms:* All eukaryotes (plants, animals, fungi, protists)<br>2. Reactants<br>• Photosynthesis<br>   o Carbon dioxide ($CO_2$)<br>   o Water ($H_2O$)<br>   o Light energy (sunlight)<br>• Cellular Respiration<br>   o Glucose ($C_6H_{12}O_6$)<br>   o Oxygen ($O_2$)<br>3. Products<br>• Photosynthesis<br>   o Glucose ($C_6H_{12}O_6$)<br>   o Oxygen ($O_2$)<br>• Cellular Respiration<br>   o Carbon dioxide ($CO_2$)<br>   o Water ($H_2O$)<br>   o ATP (energy)<br>4. Energy Flow<br>• Photosynthesis<br>   o *Energy input:* Light energy from the Sun<br>   o *Energy output:* Chemical energy stored in glucose<br>• Cellular Respiration<br>   o *Energy input:* Chemical energy from glucose<br>   o *Energy output:* Usable energy in the form of ATP<br>5. Organelles Involved<br>• Photosynthesis<br>   o Chloroplast<br>      ▪ Thylakoids (light reactions)<br>      ▪ Stroma (Calvin cycle)<br>• Cellular Respiration<br>   o Mitochondrion<br>      ▪ Matrix (Krebs cycle)<br>      ▪ Inner membrane (Electron transport chain)<br>6. Ecological Roles<br>• Photosynthesis<br>   o Primary producers (base of food chains)<br>   o Removes $CO_2$ from the atmosphere |
| |      ○ Source of atmospheric $O_2$<br>   • Cellular Respiration<br>     ○ Consumers and decomposers use glucose to obtain energy<br>     ○ Returns $CO_2$ to the atmosphere<br>     ○ Uses $O_2$ produced by photosynthesis<br>  7. Relationship Between Processes<br>   • Interdependence<br>     ○ Products of photosynthesis are reactants for cellular respiration and vice versa<br>     ○ Together form a biological cycle (carbon and energy cycles) |
| 2.Use the concept **6.Ecological Roles** to formulate questions with a real-world scenario that require high-order thinking and provide correct one-phrase answer. | **MS Copilot**<br>Question 1: Marine biologists observe a decline in phytoplankton populations due to ocean warming. This change is expected to disrupt global carbon cycles and oxygen availability. Which ecological role of phytoplankton is most directly affected by this decline?<br>Answer: Primary production<br><br>Question 2: A city plans to build a large urban park to combat rising carbon dioxide levels and improve air quality. The design includes planting a dense forested area and installing composting systems for organic waste. Using your understanding of photosynthesis and cellular respiration, explain how this plan could help balance atmospheric gases and support the ecosystem. What is the key ecological process that connects both strategies?<br>Answer: Carbon cycling.<br><br>Question 3: After a large forest fire, carbon dioxide levels in the region spike, even though plant regrowth begins quickly. Which process occurring during decomposition of burned biomass contributes most to this $CO_2$ increase?<br>Answer: Cellular respiration<br><br>Question 4: A farmer adopts cover cropping and agroforestry to improve soil health and reduce greenhouse gas emissions. Which ecological function of these plants helps achieve lower atmospheric $CO_2$ levels?<br>Answer: Carbon fixation<br><br>Question 5: Engineers designing a life-support system for a Mars habitat include algae tanks and composting units. Which interconnected processes ensure continuous oxygen supply and carbon recycling in this closed system?<br>Answer: Photosynthesis and respiration coupling |
| 3. Present the **Question2** as an MCQ with 5 answer options. It should meet 9 criteria* | **MS Copilot**<br>A municipal environmental planning team is designing a new urban park to help mitigate rising atmospheric carbon dioxide levels and support local biodiversity. The plan includes planting a dense variety of native trees and shrubs, as well as installing composting systems to process organic waste from nearby residential areas. The team aims to enhance the park's role in regulating atmospheric gases and sustaining ecological energy flow. Based on your understanding of ecosystem processes, which key biological mechanism connects both the vegetation and composting components of this plan in maintaining atmospheric gas balance?<br>What is the primary ecological process involved?<br>A) Nitrogen fixation in root nodules<br>B) Carbon cycling through biotic processes (correct)<br>C) Water uptake by plant root systems<br>D) Pollination by local insect populations<br>E) Mineral weathering in urban soils<br><br>**ChatGPT**<br>A city government proposes building a large urban park as part of a climate resilience strategy. The park will feature a densely planted forested area to absorb carbon dioxide and several composting stations to process biodegradable municipal waste. City planners aim to reduce greenhouse gas levels and enhance local ecological functions. Based on your knowledge of photosynthesis, cellular respiration, and nutrient cycles, what is the key ecological process that integrates both carbon sequestration by trees and organic matter breakdown by decomposers? |

|   | A) Oxygen transport through vascular plant tissues<br>B) Biomass accumulation in autotrophic organisms<br>C) Water cycling between soil and atmosphere<br>D) Carbon cycling through biological systems (correct)<br>E) Nitrogen fixation by root-associated microbes<br><br>**DeepAI**<br>A city plans to develop an urban park featuring dense tree planting and composting systems for organic waste, aiming enhance air quality in response to climate change concerns. The planting of trees facilitates photosynthesis, while composting accelerates organic matter decomposition, enriching soil nutrients and promoting plant growth. Considering these activities, which ecological process primarily links both?<br>A) Photosynthesis and respiration balance<br>B) Carbon cycling within ecosystems (correct)<br>C) Nitrogen fixation and mineralization<br>D) Water cycle and transpiration<br>E) Energy flow through food webs<br><br>**Wrizzle**<br>As cities grapple with deteriorating air quality, one city has proposed the construction of a large urban park. This ambitious plan includes two key strategies: planting a dense forested area designed to enhance photosynthesis and installing composting systems intended to reduce organic waste. Through photosynthesis, plants convert carbon dioxide into oxygen, while composting helps recover nutrients from organic materials. By integrating both strategies, the city aims to address atmospheric gas imbalances and foster a supportive ecosystem. What is the key ecological process that connects these two strategies?<br>A) Nutrient cycling<br>B) Carbon cycling (correct)<br>C) Oxygen cycling<br>D) Water cycling<br>E) Heat cycling |

## * 9 criteria:

1. All information presented in the MCQ — including the stem, correct answer, and distractors — must be fully scientifically accurate (every phrase separately and entire text) and can be supported by references, such as scientific articles, books, or textbooks, It is unambiguous, and free from misleading simplifications, assumptions, or comparative claims that could distort understanding.
This includes:
- Avoiding oversimplified generalizations (e.g., implying that only one group possesses a function that is actually shared).
- Ensuring that comparative or superlative terms (e.g., "better," "more," "only") are used only when clearly supported by evidence and contextually valid.
- Verifying that the correct answer is not just more correct than the others, but entirely correct on its own merits and in alignment with scientific consensus.

2. The MCQ must have:
- a long stem containing at least 3 sentences,
- 5 short answer options each containing no more than 7 words.

3. The stem of the MCQ presents a realistic, context-rich scenario that exemplifies a specific, plausible instance of a broader scientific principle or pattern (for example, subject-matter pattern: 'bacteriophages can selectively kill bacteria', real-world scenario: a case of phage therapy of a certain bacterial infection, with necessary realistic details). Scenario must describe a plausible situation that could realistically occur in the relevant scientific or professional context. All actions, settings, and roles must be scientifically appropriate and consistent with how such work is typically conducted in real life or clearly framed as a model or simulation if not.

4. The entire text of the MCQ must provide a complete and self-contained set of information that allows students of a defined speciality and level of education to make a conscious decision about the correct answer. All concepts, terms, and procedures referenced in the stem must be clearly stated and should align with the expected knowledge base of the target learners (for example, the MCQ for an introductory biology course can be based on a medical-related scenario, but it shouldn't refer to specific medical knowledge). The question should not rely on inference, unstated assumptions, or ambiguous phrasing that could confuse students or require knowledge beyond the course scope.

5. There must be a clear and necessary logical connection between the stem and the question. The stem must provide specific information—such as data, context, or observations—that is essential for selecting the correct answer. It should be impossible to answer the question correctly using prior knowledge alone, without analyzing the details provided in the stem. The stem shouldn't merely introduce a topic, it must contribute necessary information to the reasoning process.

6. MCQ requires high-order thinking to select the correct answer, not just recall some fact. (Not like in this example: 'In the context of antibiotic resistance, if a hospital-acquired infection presents with a mutation arising through non-replicative transposition of a transposon, what key feature distinguishes this mechanism from replicative transposition?
- A) The transposon remains in the original location after the process (correct answer)
- B) The process involves the replication of DNA at the insertion site.
- C) There is an integration of mobile genetic elements without any DNA breakage.
- D) It creates a copy of the transposon at a new site without altering the original.
- E) The transposase enzyme is not involved in the mechanism.')

7. Only one answer option must be clearly and definitively correct based on the information provided in the stem. The stem must include sufficient detail to rule out all other options as incorrect, incomplete, or less appropriate. If more than one option could reasonably apply without additional clarification, the criterion is not met.

8. All distractors must be conceptually related to the same domain or framework as the correct answer and be plausible responses based on the scenario. They should require a comparable level of reasoning and appear equally relevant or attractive to a student who does not yet know the correct answer. There should be no distractors that are clearly irrelevant, implausible, or based on concepts not supported or implied by the information in the stem.

9. The stem and the correct answer should not share exact or closely related key terms—especially those central to the question's logic—unless those terms are also represented in one or more distractors. A term is considered "closely related" if:
- Exact word matches, including abbreviations, acronyms, and sharing the same root or meaning (e.g., "adhesins" vs "adhesion factors"),
- Synonyms or paraphrases that provide semantic and conceptual overlap, even if wording differ (e.g., "biofilm formation" vs "quorum sensing-dependent attachment"),
- It would be recognized by a test-wise student as semantically linked based on the context provided in the stem.

Shared general terms (e.g., "cell," "bacteria") are acceptable, but repetition of unique terminology or context-specific words (e.g., "nutrient," "porin," "mutation", "resistance") should be avoided if it makes the correct answer more salient than the others.

Appendix B

**Example of using Tversky's Feature Model for calculation of contextual and linguistic similarity of MCQs within series.**

**Used AI**: MS Copilot
**Topic:** Immunity. Biology course, upper secondary school.

**Prompt 1:** In the file uploaded, identify the contextual features of each MCQ. 'Context' means the concrete manifestations of *Herd immunity*. This could relate to human health, livestock, food safety, environment, education, fundamental research in different areas and so on, as well as specific locations or participants.

Table B.1. Contextual features of MCQs revealed in response to Prompt 1.

| MCQs | Contextual features |
|---|---|
| 1(prototype). During a measles outbreak in a community with declining vaccination rates, public health officials implemented a targeted immunization campaign. Within weeks, the number of new cases dropped sharply, including among unvaccinated individuals. Which concept best explains this indirect protection effect?<br>A. Vaccination eliminates viral reservoirs<br>B. Herd immunity limits virus transmission (correct)<br>C. Vaccines increase host resistance permanently<br>D. Only children need vaccines for control<br>E. Vaccine boosters prevent all mutations' | • Measles<br>• Human health<br>• General population<br>• Public health campaign<br>• Community-level outbreak<br>• Indirect protection |
| 2. To stop a rabies outbreak, scientists placed vaccine-filled food around a forest to immunize raccoons. After a few months, there were fewer cases of rabies not only in raccoons but also in nearby animals like skunks, which were not vaccinated. What best explains why unvaccinated animals were protected?<br>A. Vaccines worked on all species<br>B. Rabies virus became less dangerous<br>C. Herd immunity stopped the virus spreading (correct)<br>D. The vaccine removed all viruses in nature<br>E. Wild animals shared antibodies through bites | • Rabies<br>• Wildlife<br>• Environmental health<br>• Raccoons, skunks<br>• Forest ecosystem<br>• Cross-species protection<br>• Environmental intervention |
| 3. A school required all students to get the whooping cough vaccine. A few months later, no new cases were reported—even among students who couldn't get vaccinated due to allergies. Health workers said the virus wasn't spreading anymore. What is the best explanation for this effect?<br>A. Sick students showed fewer symptoms<br>B. Herd immunity reduced the disease risk (correct)<br>C. The vaccine changed students' genes<br>D. Antibodies were passed by skin contact<br>E. Everyone got infected without knowing | • Whooping cough<br>• Education<br>• Human health<br>• Schoolchildren<br>• School setting<br>• Policy-based immunization<br>• Indirect protection |
| 4. In a class project, students used a computer model to study how flu spreads. When most people in the model were vaccinated, fewer people got sick—especially when more than 80% were protected. Even those without the vaccine stayed healthy.<br>Which idea does this model show?<br>A. Flu viruses become weaker over time<br>B. Herd immunity protects entire groups (correct)<br>C. Vaccines work for every type of flu<br>D. Only vaccinated people stop spreading flu<br>E. Only sick people spread viruses | • Flu<br>• Education<br>• Simulated population<br>• Computational modelling<br>• Fundamental research<br>• Threshold effects<br>• Indirect protection |
| 5. A new virus spread in a small town. Many people quickly got vaccinated, but some couldn't because of allergies. Soon after, the number of infections dropped | • New virus<br>• Human health |
| sharply in the whole town. Scientists said the virus couldn't spread easily anymore. What most likely caused this community protection?<br>A. The virus stopped spreading in air<br>B. Vaccines reduced symptoms, not infection<br>C. Herd immunity slowed virus spread (correct)<br>D. Everyone had the virus without symptoms<br>E. The virus couldn't survive in dry weather | • Town residents<br>• Small-town outbreak<br>• Rapid public health response<br>• Indirect protection |
| 6. In a community experiencing a measles outbreak, health officials launched a vaccination campaign targeting unvaccinated individuals, resulting in a significant decline in new cases, including among those unvaccinated. This outcome is best explained by which immunological principle?<br>A. Vaccines create permanent immunity in all hosts<br>B. Herd immunity reduces overall virus transmission (correct)<br>C. Vaccination directly eradicates all viral reservoirs<br>D. Only children benefit from herd immunity effects<br>E. Vaccines prevent all possible mutations in the virus | • Measles<br>• Human health<br>• Community members<br>• Public health intervention<br>• Outbreak control<br>• Indirect protection |
| 7. After a polio resurgence in a region, public health officials launched an emergency vaccination program targeting young children. Remarkably, the incidence of polio cases began to decline among not just the vaccinated children but also among their peers who remained unvaccinated. What concept clarifies this unexpected outcome?<br>A. Individual immunity guarantees disease control<br>B. Herd immunity fosters community-wide protection (correct)<br>C. Only vaccinated individuals contribute to safety<br>D. Vaccines provide immediate, lifelong immunity<br>E. Outbreaks primarily affect the unvaccinated alone | • Polio<br>• Human health<br>• Children<br>• Emergency vaccination<br>• Regional outbreak<br>• Indirect protection |
| 8. A measles outbreak occurs in a region with declining vaccination rates, prompting health officials to increase immunizations. Soon, new cases decline sharply, including among unvaccinated individuals. Which epidemiological concept explains this pattern?<br>A. Vaccines create lifelong immunity in vaccinated individuals<br>B. Herd immunity limits transmission within the community (correct)<br>C. Vaccinations prevent virus mutation entirely<br>D. Only children contribute to herd immunity effects<br>E. Vaccines eliminate the viral reservoir in the environment | • Measles<br>• Human health<br>• Regional population<br>• Vaccination coverage<br>• Regional public health<br>• Indirect protection |
| 9. Following a meningitis outbreak, a local health department expanded its immunization efforts, especially among college students. Within weeks, the overall incidence of new cases diminished, impacting both vaccinated and unvaccinated schoolmates. Which explanation best describes the dynamics at play?<br>A. Vaccines ensure total eradication of pathogens<br>B. Herd immunity lowers disease spread effectively (correct)<br>C. All individuals require vaccination to be safe<br>D. Immunization results in permanent immunity<br>E. Only severe cases warrant vaccination efforts | • Meningitis<br>• Education<br>• Human health<br>• College students<br>• Higher education setting<br>• Targeted immunization<br>• Young adult population |
| 10. Following a measles outbreak, a vaccination campaign is carried out in an under-immunized community, leading to a rapid decrease in cases, including among unvaccinated persons. What concept best explains this indirect protection?<br>A. Vaccination removes viral reservoirs entirely<br>B. Herd immunity reduces overall transmission (correct)<br>C. Vaccines increase individual resistance permanently<br>D. Only young children benefit from vaccination effects<br>E. Vaccines block all possible virus mutations | • Measles<br>• Human health<br>• Community members<br>• Under-immunized population<br>• Vaccination campaign<br>• Indirect protection |

**Prompt 2**: Here are the contexts of MCQs 1-10 (*table above*)
Calculate contextual similarity comparing MCQs 1-10 in pairs.
Use Tversky's Feature Model:
$S(A, B) = \theta f(A \cap B) - \alpha f(A-B) - \beta f(B-A)$.
Weighing values are: $\theta = 1$, $\alpha = \beta = 0.5$.

Table B.2. Contextual similarity of MCQs calculated in response to Prompt 2.

|  | MCQ 1 | MCQ 2 | MCQ 3 | MCQ 4 | MCQ 5 | MCQ 6 | MCQ 7 | MCQ 8 | MCQ 9 | MCQ 10 |
|---|---|---|---|---|---|---|---|---|---|---|
| MCQ 1 | 6 | -6,5 | -4,5 | -6,5 | -2 | 0 | -2 | 0 | -6,5 | 0 |
| MCQ 2 | -6,5 | 7 | -7 | -7 | -6,5 | -6,5 | -6,5 | -6,5 | -7 | -6,5 |
| MCQ 3 | -4,5 | -7 | 7 | -5 | -4,5 | -4,5 | -4,5 | -4,5 | -3 | -4,5 |
| MCQ 4 | -6,5 | -7 | -5 | 7 | -6,5 | -6,5 | -6,5 | -6,5 | -5 | -6,5 |
| MCQ 5 | -2 | -6,5 | -4,5 | -6,5 | 6 | -2 | -2 | -2 | -6,5 | -2 |
| MCQ 6 | 0 | -6,5 | -4,5 | -6,5 | -2 | 6 | -2 | 0 | -6,5 | 2 |
| MCQ 7 | -2 | -6,5 | -4,5 | -6,5 | -2 | -2 | 6 | -2 | -6,5 | -2 |
| MCQ 8 | 0 | -6,5 | -4,5 | -6,5 | -2 | 0 | -2 | 6 | -6,5 | 0 |
| MCQ 9 | -6,5 | -7 | -3 | -5 | -6,5 | -6,5 | -6,5 | -6,5 | 7 | -6,5 |
| MCQ 10 | 0 | -6,5 | -4,5 | -6,5 | -2 | 2 | -2 | 0 | -6,5 | 6 |

**Prompt 3:** Calculate the linguistic similarity of MCQs 1-10 in pairs.
Use Tversky's Feature Model:
$S(A, B) = \theta f(A \cap B) - \alpha f(A - B) - \beta f(B - A)$.
Weighing values are: $\theta = 1$, $\alpha = \beta = 0.5$.

Table B.3. Linguistic similarity of MCQs calculated in response to Prompt 3.

|  | MCQ 1 | MCQ 2 | MCQ 3 | MCQ 4 | MCQ 5 | MCQ 6 | MCQ 7 | MCQ 8 | MCQ 9 | MCQ 10 |
|---|---|---|---|---|---|---|---|---|---|---|
| MCQ 1 | 65.0 | -23.0 | -27.5 | -31.5 | -24.0 | 8.0 | -14.5 | 16.0 | -13.0 | 8.0 |
| MCQ 2 | -23.0 | 65.0 | -21.5 | -19.5 | -22.0 | -26.0 | -24.5 | -24.0 | -25.0 | -24.0 |
| MCQ 3 | -27.5 | -21.5 | 66.0 | -22.0 | -14.5 | -26.5 | -33.0 | -30.5 | -27.5 | -30.5 |
| MCQ 4 | -31.5 | -19.5 | -22.0 | 62.0 | -22.5 | -32.5 | -35.0 | -30.5 | -27.5 | -36.5 |
| MCQ 5 | -24.0 | -22.0 | -14.5 | -22.5 | 59.0 | -31.0 | -25.5 | -25.0 | -34.0 | -31.0 |
| MCQ 6 | 8.0 | -26.0 | -26.5 | -32.5 | -31.0 | 59.0 | -13.5 | 7.0 | -14.0 | 11.0 |
| MCQ 7 | -14.5 | -24.5 | -33.0 | -35.0 | -25.5 | -13.5 | 68.0 | -7.5 | -22.5 | -15.5 |
| MCQ 8 | 16.0 | -24.0 | -30.5 | -30.5 | -25.0 | 7.0 | -7.5 | 59.0 | -16.0 | -1.0 |
| MCQ 9 | -13.0 | -25.0 | -27.5 | -27.5 | -34.0 | -14.0 | -22.5 | -16.0 | 65.0 | -22.0 |
| MCQ 10 | 8.0 | -24.0 | -30.5 | -36.5 | -31.0 | 11.0 | -15.5 | -1.0 | -22.0 | 59.0 |